# Market Depth and Risk Return Analysis of Dhaka Stock Exchange: An Empirical Test of Market Efficiency

**Md. Mahmudul Alam**
Officer, Customer Relationship Management (CRM)
Brands, Commercial Division
Grameenphone Ltd.
Dhaka, Bangladesh
Email: rony000@gmail.com

**Kazi Ashraful Alam**
Assistant Professor
Faculty of Business
ASA University Bangladesh
Dhaka, Bangladesh
Email: Kashraf@presidency.edu.bd

**Md. Gazi Salah Uddin**
Lecturer
School of Business
Presidency University
Dhaka, Bangladesh
Email: salahuddin@presidency.edu.bd

# Citation Reference:

# Market Depth and Risk Return Analysis of Dhaka Stock Exchange: An Empirical Test of Market Efficiency

**Abstract**
It is customary that when security prices fully reflect all available information, the markets for those securities are said to be efficient. And if markets are inefficient, investors can use available information ignored by the market to earn abnormally high returns on their investments. In this context this paper tries to find evidence supporting the reality of weak-form efficiency of the Dhaka Stock Exchange (DSE) by examining the issues of market risk-return relationship and market depth or liquidity for DSE. The study uses a data set of daily market index and returns for the period of 1994 to 2005 and weekly market capital turnover in proportion of total market capital for the period of 1994 to 2005. The paper also looks about the market risk (systemic risk) and return where it is found that market rate of return of DSE is very low or sometimes negative. Eventually Capital Asset Pricing Model (CAPM), which envisages the relationship between risk and the expected rate of return on a risky security, is found unrelated in DSE market. As proper risk-return relationships of the market is seems to be deficient in DSE and the market is not liquid, interest of the available investors are bring into being very insignificant. All these issues are very noteworthy to the security analysts, investors and security exchange regulatory bodies in their policy making decisions to progress the market condition.

## Introduction

Emerging markets are typically characterized by low liquidity, unreliable information and considerable volatility. If there is any wrong information or the flow of information is not rapid, or the price is not adjusted to the information, then that market would cease to be considered as an inefficient market. Good investors always look for investing in an efficient market, therefore a significant number of buyers and sellers are available in an efficient market and the overall market transaction turns very high.

The Efficient Market Hypothesis (EMH) and the empirical tests of the hypothesis can be divided based on the information set involved: Weak-form EMH, Semi-strong form EMH and Strong-form EMH. Weak-form market efficiency hypothesis (EMH) states that the stock returns are serially un-correlated and have a constant mean. In other words, a market is considered weak form efficient if current prices fully reflect all information contained in historical prices, which implies that no investor can devise a trading rule based solely on past price patterns to earn abnormal returns. Whereas, the semi- strong form EMH means that securities prices adjust rapidly based on all public information and the strong-form EMH contends that stock prices fully reflect all private and public information.

There are two types of risk in capital market investment- uncontrollable or systematic risk (non-diversifiable) and controllable or unsystematic risk (diversifiable). When determining the risk of a share, investor only looks on controllable risk that can be measured by Capital Assets Pricing Model (CAPM) model, which is one of the widely used methods of assets valuation. If the market returns gives wrong indication CAPM model will not work perfectly. This indicates high inefficiency of the market.

One of the important efficiency indicators of the market is liquidity or Market Depth. Market Depth means numerous buyers and sellers are available in the market. As a result potential buyers and sellers are easily found at any time for transaction of any company's share. For that reason in the market radical price change is prevented. It can be measured by observing transaction values proportion out of total market capital. If this proportion is high, then the market indicates its depth is also high.

Several relevant studies revealed that the markets in developing and less developed countries are not efficient in semi-strong form or strong form. The test of semi strong form and strong form efficiency is found insignificant in less developed countries. So, the study investigates to find out the facts whether the Dhaka Stock Exchange follows random walk model or the market is weak form efficient. The researcher also tried to find out the Market Risk Return relationship and the frequency of Market Depth on Dhaka Stock Exchange.

## Literature Review
Early findings on market efficiency differ among researchers. Fama (1965), Samuelson (1965), and Working (1960) used random walk model and found market is efficient. Shiller (1989) shown that stock prices do follow a random walk and also explained the reason behind the behavior of the stock prices. Niederhoffer & Osborne (1966) found an early rejection of a random walk model. Poterba & Summers (1988) argued that there is little theoretical basis for strong attachment to the null hypothesis that stock prices follow a random walk. Lo & MacKinlay (1988) investigated the sampling distributions of variance ratios over different sampling intervals and found that stock returns do not follow a random walk. Claessens (1995) in a world bank study reported significant serial correlation in equity returns from 19 emerging markets and suggested that stock prices in emerging markets violates weak form EMH. Poshakwale (1996) found the evidence of non-randomness stock price behavior and the market inefficiency (not weak-form efficient) on the Indian market. Khababa (1998) has examined the behavior of stock price in the Saudi Financial market seeking the evidence for weak-form efficiency and found that the market was not weak-form efficient. He explained that the inefficiency might be due to delay in operations and high transaction cost, thinness of trading and illiquidity in the market.

Surprisingly researchers also find different types of results while working on the same market. Jammine and Hawkins (1974), Hadassin (1976) and Du Toit (1986) reject weak-form efficiency, but Affleck-Graves and Money (1975), Gilbertson and Roux (1977, 1978) found weak-form efficiency; Knight and Afflect-Graves (1983) rejected semi-strong form efficiency but Knight, Afflect-Graves and Hamman (1985) showed semi-strong form efficiency; Gilbertson (1976) found evidence supporting strong-form efficiency, but Knight and Firer (1989) rejected the strong-form efficiency on Johannesburg Stock Exchange (JSE). Harvey (1993) stated that stock returns of emerging countries are highly predictable and have low correlation with stock returns of developed countries. He concluded that emerging markets are less efficient than developed markets and that higher return and low risk can be obtained by incorporating emerging market stocks in investors' portfolios. Balaban 1995, Urrutia 1995, Grieb & Reyes 1999, Kawakatsu & Morey 1999 showed non randomness of stock prices for emerging markets. Few studies have already been conducted on Dhaka Stock Exchange (DSE). Hassan (1999) studied on time-varying risk return relationship for Bangladesh by utilizing a unique data set of daily stock prices and returns. The result found that DSE equity returns held positive skewness, excess kurtosis and deviation from normality and the returns displayed significant serial correlation, implying the stock market is inefficient. Mobarek (2000) concluded that Dhaka Stock Exchange does not follow random walk model and there are significant autocorrelation causes to DSE is not weak form efficient. Their result did not change for different sub-sample observations, without outlier, and for individual securities. Haque (2001) worked on the cumulative abnormal profit on the study period. He described the experience of DSE after the scam of November 1996 by applying CAPM and EMH. Based on the data four months before and four months after the automation, the paper measured risk-return performance, estimated SML for big capital and small capital companies before and after automation and tested EMH. The test results indicated that the market does not improve, and even after automation manipulation continued.

Kader (2005) has no evidence that Dhaka Stock Exchange is weak form efficient by testing whether any technical trading strategy yielded abnormal profit or not by using technical trading rule (K% filter rule). Islam (2005) analyzed on the predictability of the share price in Dhaka Stock Exchange prior to the boom in 1996 and by using heteroscedasticity-robust tests found evidence in favor of short-term predictability of share prices in the Dhaka stock market prior to the 1996 boom, but not during the post-crash period. After thorough investigation it was concluded that the Securities and Exchange Commission were able to give more transparency of the Dhaka Stock Exchange by taking various steps. Uddin and Alam (2007) examines the linear relationship between share price and interest rate, share price and growth of interest rate, growth of share price and interest rate, and growth of share price and growth of interest rate were determined through ordinary least-square (OLS) regression. For all of the cases, included and excluded outlier, they found that Interest Rate has significant negative relationship with Share Price and Growth of Interest Rate has significant negative relationship with Growth of Share Price in Dhaka Stock Market so that DSE is not weak form efficient.

## Data, Model and Methodology

The study uses a data set of 3,209 daily observations on market index and returns for the period of 1994 to 2005 and 3,057 weekly observations on market capital turnover in proportion of total market capital for the period of 1994 to 2005. These data are collected from Dhaka Stock Exchange.

To test the EMH (Efficient Market Hypothesis) of DSE, the tools of stationarity (Unit Root Test) of share prices is tested by using daily market returns. DSE prepares daily price index from daily weighted-average price of daily transaction of each stock. Daily market returns ($R_t$) are calculated from the daily price indices such as follows:

$$R_t = Ln(PI_t / PI_{t-1}) \qquad (1)$$

Where,
$R_t$ = market return at period t;
$PI_t$ = price index at period t;
$PI_{t-1}$ = the price index at period t-1 and
ln = natural log.

While calculating the market return (Eq-1) for the efficiency test logarithm is used because it is justified by both theoretically and empirically. Theoretically, logarithmic returns are analytically more tractable when linking returns over longer intervals. Empirically, logarithmic returns are more likely to be normally distributed which is a prior condition of standard statistical techniques (Strong 1992).

CAPM model can be viewed both as a mathematical equation and graphically as Security Market Line (SML) which shows the relationship between risk and return. The significance of CAPM model is analyzed for determining expected returns in respect of risk of an investor.

$$\textbf{CAPM Model:} \quad K_j = K_{rf} + \beta_j (K_m - K_{rf}) \qquad (2)$$

Here,
$K_j$ = Return for the jth Security,
$K_{rf}$ = Risk Free Rate
$\beta_j$ = Individual Company Risk,
$K_m$ = Expected Return for the market

To determine the market depth for DSE, descriptive statistics and the ratio of Market Capital to Market Turnover is analyzed for the mentioned period.

## Empirical Result and Discussion

**Randomness of market return**
The statistical output of unit root test for market return series suggests that there are serial dependencies of return of Dhaka Stock Exchange. ADF calculated values are significant at 1% level for all 10 degrees of freedom (lags) suggests that the return series does not follow random walk model (Table-1) that means DSE is not efficient in weak form.

Table 1: Output for Unit Root Test on DSE Return (for 10 lag period) (Eq-1)

| Lag Year | ADF Calculated Value | ADF Critical Value at 1% | ADF Critical Value at 5% |
|---|---|---|---|
| 1 | -37.48 | -3.923 | -3.066 |
| 2 | -28.89 | -3.923 | -3.066 |

| 3  | -24.55 | -3.923 | -3.066 |
| 4  | -22.02 | -3.923 | -3.066 |
| 5  | -21.04 | -3.923 | -3.066 |
| 6  | -19.27 | -3.923 | -3.066 |
| 7  | -18.02 | -3.923 | -3.066 |
| 8  | -17.00 | -3.923 | -3.066 |
| 9  | -15.23 | -3.923 | -3.066 |
| 10 | -14.20 | -3.923 | -3.066 |

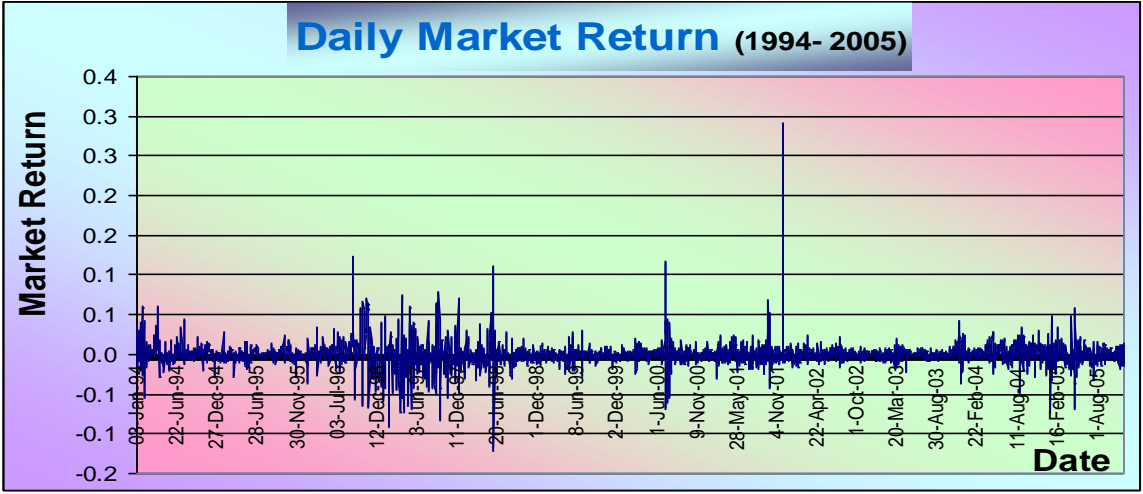

**Market risk-return relationships**

The DSE market indicates an opposite direction from rationality. Long term observations of twelve years based on daily capital return from the market shows the index is very low; on an average daily basis return is 0.0442% and for short term sometimes it was negative. While excluding the outlier, based on 3057 observations, the return turns lower to 0.0139%. This may cause a rational investor to be less interested to invest in this market.

**Table 2: Average Market return of DSE**

| Time Range | Average Market Return | Return As % Term |
| --- | --- | --- |
| 1/1/1994 - 31/12/2005 | 0.000442 | 0.0442% |
| 1/1/1994 - 31/12/2005 (Excluding Outlier) | .000139 | 0.0139% |

If the market return gives wrong indication (e.g. very low or negative return) CAPM model never works perfectly. Beta measures non-diversifiable or systematic risk and the beta for the market is 1. In such cases, when market provides very low or negative return, expected returns of high risky company will become negative. Bangladesh government T-bill rate ($K_{rf}$) is 5%, DSE market return is .0442%, and suppose 'j' company's beta is 1.5. For this 'j' company, expected return will be {5+1.5(.0442-4.5)} = -1.68%. Thus, when market provides negative return ($K_m$), CAPM model provides negative expected return for high risky company, which is irrational. A company with a high risk should provide comparatively a high return, but for the reason of a low market return CAPM model shows wrong expected return for that company. This indicates DSE is an inefficient market.

**Frequency of the market depth**
From the weekly observations of twelve years on DSE, average daily transaction (BDT) to total market capital is only 0.1539%. On the other hand, as one of the most efficient markets, NYSE average daily capital transaction of total market capital is over 100%. Though NYSE is considered as a model of efficient market and it is totally different from DSE, NYSE's daily capital transaction of its total market capital is more than 500 times higher of DSE. This very low amount of turnover indicates DSE market is not liquid enough. As a result, a significant number of buyers and sellers are not available in the DSE and active investors in DSE are very few.

Table 3: Average Market Capital to Turnover Ratio of DSE

| Time Range | Market Capital (Cr. Tk) | Turnover Value (Cr. Tk) | Turnover by Market Capital (Weekly Basis) | Standard Deviation |
|---|---|---|---|---|
| 1994 | 3126.59 | 1.7 | .057% | .0009 |
| 1995 | 4627.36 | 2.24 | .0462% | .0003 |
| 1996 | 9851.53 | 9.89 | .098% | .0006 |
| 1997 | 9377.96 | 6.69 | .075% | .0008 |
| 1998 | 5679.08 | 13.56 | .2537% | .0031 |
| 1999 | 4,691.97 | 15.31 | 0.3244% | .0014 |
| 2000 | 5,399.90 | 13.62 | 0.240% | .0013 |
| 2001 | 6,349.34 | 14.99 | 0.2331% | .001 |
| 2002 | 6,486.93 | 11.31 | 0.1724% | .0006 |
| 2003 | 6,859.27 | 6.89 | 0.0985% | .0007 |
| 2004 | 14,117.62 | 19.44 | 0.127% | .0006 |
| 2005 | 21,904.33 | 24.31 | 0.11% | .0004 |
| 1/1/1994 - 31/12/2005 | 8266.114 | 11.79 | 0.1539% | 0.0015 |

High Standard Deviation means- daily capital transaction of total market capital is fluctuating more from its mean (average). It shows an indication that there may be enough passive investors in that market, and when passive investors execute their equity then market turnover will be high and the market would fluctuate more. But the standard deviation of daily capital transaction of total market capital is 0.00147 in weekly average basis, indicating DSE's passive investors is also very small. So, both active and passive types of investors are highly absent in DSE.

Table 4: Descriptive Statistics for the Variables of Market Depth Measurement of DSE

| | Minimum | Maximum | Mean | Std. Deviation |
|---|---|---|---|---|
| Market Capital (Cr. TK) | 1,522.78 | 25,841.940 | 8,266.114 | 5,504.601 |
| Turnover Value (Cr. TK) | 0.1230000 | 64.73000 | 11.79024 | 10.60712 |
| Market Turnover / Market Capital (Weekly Basis) | 0.0000273 | 0.01252 | 0.00154 | 0.00147 |

In risk-return analysis it is mentioned that when the market return is negative or very low, no rational investor will invest at that market. Rational investors will switch to another market where they can get more return bearing same risk level. In Bangladesh the interest rate is very high (government T-bill 4.5%, savings account interest 6%) in proportion to share market risk free return (average 0.0442%). So investors keep their money in the bank or invest in other sectors rather than investing in the stock market. This is the reason why the commercial banks hold more than ten thousand crores taka idle money at present. As a consequence, significant buyers and sellers are not available in the market and the market liquidity or market depth is very low. As a result, potential buyers and sellers are not found at any time for transaction of any company's share, and dramatic price change cannot be prevented. Ultimately DSE is also found inefficient for lacking of enough liquidity (market depth).

## Conclusion

The significance of this research is to find out the efficiency of the Dhaka Stock Exchange (DSE) through analyzing the randomness of market return, market risk-return relationships and the frequency of the market depth or liquidity. To the best of the researcher's knowledge, this is a new study of measuring market efficiency through market depth and risk- return relationship. The paper found that DSE is not efficient, and the CAPM is not working in this market. Under this circumstance, a new model of asset valuation is highly recommended to develop for the emerging market. In Bangladesh the interest rate is very high in respect of capital market return so that investors keep their money in the bank or invest in other sectors rather than investing in the stock market. As a result, available traders are not found at any time for transactions.

This study is useful for a number of reasons. The results of this study will be of great interest to academics, policy makers and local and foreign listed and unlisted companies. Moreover, globalization in the world economy has created an enormous opportunity for the investors to diversify their portfolios across the globe. As a result, this study examining the efficiency and other characteristics of DSE markets would be of great benefit to investors at home and abroad. Finally, it may also be useful for international organizations (such as the World Bank, IMF, WTO) and governments of development partners who are interested in the development of capital markets in the emerging countries.

## References

Affleck-Graves, J. F. and A. H. Money. (1975). 'A Note on the Random Walk Model and South African Share Prices', *The South African Journal of Economics*, Vol. 43, No. 3, pp. 382-388.
Balaban, E. (1995). 'Day of the Week Effects: New Evidence from an Emerging Stock Market', *Applied Economics Letters*, Vol. 2, No. 5, pp. 139-43.
Claessens, S., S. Dasgupta and J. Glen. (1995). 'Return Behaviour in Emerging Stock Markets', *The World Bank Economic Review*, vol.9 (1), pp.131–151
Du Toit, G. S. (1986). 'Technical Analysis and Market Efficiency on the Johannesburg Stock Exchange', *Working Paper for D.Com degree, Pretoria*: University of Pretoria.

Fama, E. (1965). 'The Behavior of Stock Market Prices', *Journal of Business*, Vol 38, pp. 34-105.

Gilbertson, B. P. (1976). 'The Performance of South African Mutual Funds', *Johannesburg: Johannesburg Consolidated Investment Company,* Report No. F76/84.

Gilbertson, B. P. and F. J. P. Roux. (1978). 'Some Further Comments on the Johannesburg Stock Exchange as an Efficient Market', *Investment Analysts Journal*, Vol. 11, pp. 21-30.

Grieb, T. A. and M. G. Reyes. (1999). 'Random Walk Tests for Latin American Equity Indexes and Individual Firms', *Journal of Financial Research*, Vol. 22, No. 4, pp. 371-383.

Hadassin, I. (1976). 'An Investigation into the Bahavior of Emerging and Share Prices of South African Listed Companies', *Investment Analysts Journal*, Vol. 8, pp. 13-24.

Haque, M. Shamsul, R. Eunus and M. Ahmed. (2001). 'Risk Return & Market Efficiency in Capital Market under Distress: Theory and Evidence from DSE', Chittagong Stock Exchange Publication, Quarter-1. Available at <http://www.csebd.com/cse/Publications/portfolio_Q1_2001/risk_%20return.htm>.

Harvey, A. C. 1993, Time Series Models, 2nd Ed., Harvester Wheatsheaf, New York

Hassan, M. Kabir, M. A. Islam and S. A. Basher. (1999), 'Market Efficiency, Time-Varying Volatility and Equity Returns in Bangladesh Stock Market', Working Papers 2002-06, York University, Department of Economics, revised Jun 2002.

Islam, Ainul and Mohammed Khaled. (2005). 'Tests Of Weak-Form Efficiency Of The Dhaka Stock Exchange', *Journal Of Business Finance & Accounting*, vol.32(7-8), pp.1613-1624, September/October.

Jammine, A. P. and D. M. Hawkins. (1974). 'The Behavior of Some Share Indices: A Statistical Analysis', *The South African Journal of Economics*, Vol. 42, No. 1, pp. 43-55.

Kader, A. A., and, A. F. M A. Rahman.( 2005), 'Testing the Weak-Form Efficiency of an Emerging Market: Evidence from the Dhaka Stock Exchange of Bangladesh', *AIUB Journal*, vol.4(2), August.

Kawakatsu, H. and M. R. Morey. (1999). 'An Empirical Examination of Financial Liberalization and the Efficiency of Emerging Market Stock Prices', *The Journal of Financial Research,* Vol. 22, pp.385-411

Khababa, Nourredine. (1998). 'Behavior of stock prices in the Saudi Arabian Financial Market: Empirical research findings', *Journal of Financial Management & Analysis,* vol.11(1), pp.48-55, Jan-June.

Knight, R. F., J. F. Affleck-Graves and W. D. Hamman. (1985). 'The Effect of Inventory Valuation Methods on Share Prices: Some New Evidence for the JSE', *Investment Analysts Journal,* Vol. 26, pp. 45-47.

Knight, R. F., J. F. Affleck-Graves.( 1983). 'The Efficient Market Hypothesis and a Change to LIFO: An Empirical Study on the JSE', *Investment Analysts Journal,* Vol. 21, pp. 21-33.

Lo, A.W. and A.C. Mackinlay. (1988). 'Stock Market Prices Do Not Follow Random Walks: Evidence from A Simple Specification Test', *Review of Financial Studies,* Vol 1, pp. 41-66.

Mobarek, Asma, and Keasey, Keavin. (2000). 'Weak-form market efficiency of an emerging market: Evidence from Dhaka Stock Market of Bangladesh', *Paper presented at the ENBS Conference held on Osio,* May 2000.

Niederhoffer, V. and M.F.M. Osborne. (1966). 'Market Making and Reversal on the Stock Exchange', *Journal of the American Statistical Association*, Vol.61, pp. 897-916.

Poshakwale, S. (1996). 'Evidence on the Weak-form efficiency and the day of the week effect in the Indian Stock Market', *Finance India*, vol.10 (3), pp.605-616, September

Poterba, J. M. and L. H. Summers.(1988). 'Mean Reversion in Stock Returns: Evidence and Implications', *Journal of Financial Economics*, Vol. 22, pp. 27-59.

Samuelson, P. A. (1965). 'Proof that Properly Anticipated Prices Fluctuate Randomly', *Industrial Management Review*, Vol. 6, pp. 41-49.

Shiller, Robert J. (1989). Market Volatility, M.I.T. Press, Mass.

Strong, N. (1992) 'Modelling Abnormal Returns: A Review Article', *Journal of Business Finance and Accounting,* vol.19 (4), pp.533-553, June.

Uddin, Md Gazi Salah and Alam, Md. Mahmudul. (2007). 'The Impacts of Interest Rate on Stock Market: Empirical Evidence from Dhaka Stock Exchange', *South Asian Journal of Management and Sciences,* Vol. 1, Issues 2. September.

Urrutia, J. L. (1995). 'Tests of Random Walk and Market Efficiency for Latin American Emerging Markets', *Journal of Financial Research*, Vol. 18, pp. 299-309